\newcommand{\be}{\begin{equation}}
\newcommand{\ee}{\end{equation}}
\newcommand{\bea}{\begin{eqnarray}}
\newcommand{\eea}{\end{eqnarray}}
\newcommand{\bfig}{\begin{figure}}
\newcommand{\efig}{\end{figure}}

\newcommand{\rb}{\mbox{$\bbox{r}$}}
\newcommand{\ett}{\mbox{$\tilde{\eta}$}}
\newcommand{\force}{\mbox{$\bbox{F}$}}
\newcommand{\fc}{\mbox{$F_{\text{c}}$}}
\newcommand{\e}{\mbox{${\cal E}$}}
\newcommand{\zf}{\mbox{$\tilde{Z}$}}

\def\mypprint{\global\let\ifmypprint\iftrue}
\global\let\iftorefs\iffalse
\def\torefs{\global\let\iftorefs\iftrue}
\global\let\dofloatfig\iffalse
\def\floatthefig{\let\dofloatfig\iftrue}
\mypprint
\ifmypprint
\documentstyle[pre,aps,multicol,epsf,rotate,twoside]{revtex}

\floatthefig
\else
  \iftorefs
    \catcode`\@=11
    
    \def\figure{\let\@capwidth\columnwidth\@float{figure}}
    \let\endfigure\end@float
    \@namedef{figure*}{\let\@capwidth\textwidth\@dblfloat{figure}}
    \@namedef{endfigure*}{\end@dblfloat}
    \catcode`\@=12
     \floatthefig
  \fi
\fi
\input epsf.tex

\begin{document}
\draft

\twocolumn[\hsize\textwidth\columnwidth\hsize\csname @twocolumnfalse\endcsname

\title{Pulling Pinned Polymers and Unzipping DNA}

\author{David K. Lubensky\cite{my-email} and
David R. Nelson\cite{nelson-email}}

\address{Department of Physics, Harvard University, Cambridge MA
02138}

\maketitle



\begin{abstract}
We study a class of micromanipulation experiments, exemplified by the
pulling apart of the two strands of double-stranded DNA (dsDNA).  When
the pulling force is increased to a critical value, an ``unzipping''
transition occurs.  For random DNA sequences with short-ranged
correlations, we obtain exact results for the number of monomers
liberated and the specific heat, including the critical
behavior at the transition.  Related systems include a random
heteropolymer pulled away from an adsorbing surface and a vortex line
in a type II superconductor tilted away from a fragmented columnar
defect.

\end{abstract}


\vskip2pc]


\par 
Recent years have seen an explosion in the use of single molecule techniques to probe biological and other ``soft''
materials.  It now possible, for example, to monitor the
breaking of individual ``lock and key''
bonds~\cite{lock-key} and the unfolding of individual
proteins~\cite{protein}; the mechanical properties of
single DNA molecules~\cite{dna-pull}
and the behavior of single molecular motors~\cite{motors} have
been characterized in great detail.  In
contrast to more traditional experiments, these new approaches give
access to fluctuations on the scale of individual molecules,
without the requirement for averaging over a macroscopic sample.
One can, moreover, now push or pull directly on a micron-sized
object, and watch how it responds.  The potentially profound
implications both for complex fluids and for biological
physics---where single molecule techniques can often more closely
mimic conditions in the cell than conventional assays---are only
beginning to be explored.

\par Despite a number of notable contributions, theory has often been
out-paced by these rapid experimental advances.  Certainly, the tools
available to analyze single-molecule experiments have not yet reached
the level of sophistication and generality of theories of mesoscopic
quantum systems.  This is especially true when it comes to the role of
quenched randomness, which, though often present, is typically
neglected in initial theories of a given system.  This Letter seeks to
fill some of this gap.  We study a class of micromanipulation
experiments in which a polymer or other line-like object is pulled
away from a confining potential well.  An example of such a situation
is the pulling apart of the two strands of dsDNA (fig.~\ref{fig1}).
Formally, the distance between the two strands may be viewed as the
coordinate of a single polymer, and the base-pairing interactions
between complementary strands as a potential well.  At a critical
value of the pulling force, a novel phase transition occurs in which
the two strands are pulled completely apart.  Aspects of this
transition for a homopolymer (or, equivalently, DNA with all base
pairs the same) have been studied in a related model of a flux line in
a type II superconductor~\cite{bhatt,hatano}.  Here, we show that the
transition is markedly different for random heteropolymers.  In
particular, the number of monomers liberated as the transition is
approached diverges much more strongly for heteropolymers than for
homopolymers; similar differences appear in the specific heat.  We
calculate {\em exact} critical exponents and crossover functions for
the random case.

\par
Figure~\ref{fig1}
sketches the DNA-opening experiment: One of the two single strands of
a dsDNA molecule is attached to a glass slide, while a constant force
\force\ directed away from the slide acts on the end of the other
strand.  Methods for exerting a constant force on the
piconewton scale have been developed by several
groups~\cite{dna-pull,const-force}.  Under the influence of the force
\force, the DNA partially ``unzips'' at one end, breaking $m$ base
pairs.  In thermal equilibrium, the degree of opening $m$ is of course
a fluctuating quantity.  Because the base sequence of protein-coding
DNA appears to many statistical tests to be random and uncorrelated
along the backbone (at least up to a length\linebreak\vspace{-10pt}
\dofloatfig
\bfig
\epsffile{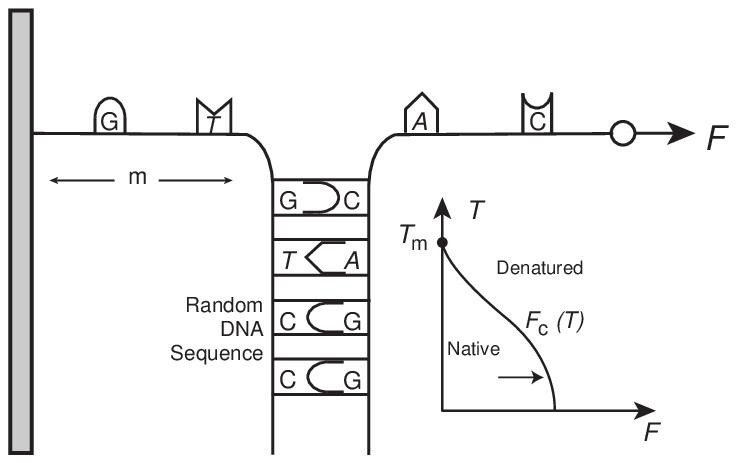}
\caption{Sketch of the ``unzipping DNA'' experiment.  One of the single strands of a dsDNA molecule with a
random base sequence is attached to a glass slide and the other is
pulled away from the slide with a constant force $F$.  As a result,
the two strands partially separate, breaking $m$ bonds ($m=2$ in the
figure). {\em Inset}: Schematic phase diagram in the
temperature--pulling force ($T$--$F$) plane for a dsDNA molecule in 3
dimensions.  At low $T$ and $F$, the polymer is in the native,
double-stranded state.  At the phase transition line $\fc(T)$, the DNA
denatures, and the two strands separate.  As indicated by the arrow,
here we consider the {\em unzipping} transition that occurs when the
transition is approached away from the $F=0$ {\em melting} temperature
$T_{\text{m}}$.  \label{fig1} }
\efig
\fi
\noindent scale set by the
sequence's mosaic structure)\cite{dna-statistics}, the free energy
landscape in which $m$ fluctuates can be taken to have a quenched
random component.  Bockelmann, Essevaz-Roulet, and Heslot have
performed an elegant series of experiments in a different statistical
ensemble, measuring the average force required to hold the positions
of both single strands fixed~\cite{heslot,siggia-heslot}.  However,
because of subtleties associated with the thermodynamic limit in a
single molecule system (see below), the two ensembles are not
equivalent.

\par
In the remainder of this paper, we first introduce a
coarse-grained model for the interaction of the two single strands of
the dsDNA.  By focusing on the {\em unzipping} transition induced by
pulling on the single strands, we can avoid treating most of the
degrees of freedom explicitly, obtaining a problem that can be solved
exactly by a mapping to a Markov process.
Although for concreteness we will focus primarily on the DNA-opening
realization of our model, our results also apply to a number of
related physical systems, some of which will be described in the
conclusion.  Throughout, we set $k_{\text{B}} =
1$.

\par 
The bulk {\em melting} transition of dsDNA (see inset to
Fig.~\ref{fig1}) can be described by a Peyrard-Bishop-like
model~\cite{hwa}.  One
views the two single strands as Gaussian polymers whose
$n^{\text{th}}$ monomers have positions $\rb_1(n)$ and $\rb_2(n)$.
Below the melting transition, it should be possible to neglect
non-native base pairings.  The interactions between the two strands,
coarse-grained over a number of bases, can then be described by a
phenomenological potential energy term $V_n[\rb(n)] = [1 + \ett(n)]
h[\rb(n)]$.  Here $h$ is a short-ranged attractive potential
whose strength is temperature-dependent, $\rb \equiv \rb_1 -
\rb_2$, and the variation with base sequence of the strength of the
attraction between strands is described by $\ett(n)$, which we take to
be a random variable with short-ranged correlations.  The effective
Hamiltonian then becomes, up to an uninteresting center of mass term,
\be 
{\cal H_{\text{melt}}} = \int_0^N dn \left\{ \, \frac{T d}{2 b^2}
\left(\frac{d \rb}{dn}\right)^2 + V_n[\rb(n)] \, \right\} 
\ee 
 where $d$ is the spatial
dimension and $b$ is $\sqrt{2}$ times the Kuhn length of single-stranded DNA.  Standard arguments show that the
partition function $Z_{\text{melt}} \equiv \int {\cal D}[\rb(n)]
\exp(-{\cal H_{\text{melt}} }/T)$ obeys an (imaginary) time-dependent Schr\"{o}dinger
equation.

\par 
The unzipping transition may be studied by adding to ${\cal
H_{\text{melt}}}$ the term $
{\cal H}_{\text{pull}} = \force \cdot \rb(0) = \force \cdot
\rb(N) - \int_0^N dn \, \force \cdot d \rb/dn$.
If the term proportional
to $\rb(N)$, which should not affect the opening near $n=0$ for a
sufficiently long polymer, is dropped, a time-dependent version of the
``non-Hermitian quantum mechanics'' studied in~\cite{hatano} results.
In the time-independent case (corresponding to pulling apart homopolymeric dsDNA) there is a sharp first order unzipping
transition at a critical value of the pulling force \fc\ satisfying
$\epsilon_0(T) = -\fc^2 b^2/(2 d T)$, where $\epsilon_0 < 0$ is the ground
state energy of the Hermitian quantum mechanics problem obtained by
setting $F = 0$.  In general, $-F^2 b^2/(2 d T)$ is the free energy per monomer
of the unzipped monomers aligned with the pulling force.  The free
energy per monomer of the dsDNA that has remained zipped is
$\epsilon_0$, independent of $F$.
The physical interpretation of the unzipping transition is thus clear: For $F < \fc(T)$, the DNA
minimizes its free energy by remaining in the double-stranded form,
while for $F > \fc$ it is advantageous to pull apart as many bases as
possible.  As $ F \rightarrow \fc^{-}$, the free energy difference between
the bound, double-stranded phase and the pulled out, single-stranded
phase becomes very small, and thermal fluctuations
unbind a large number of monomers near the end of the DNA.  As the transition is
approached, the equilibrium ensemble average of the number $m$ of
monomers that are pulled out diverges like
\be
\langle m \rangle \sim (\fc -F)^{-1} \, . \;
\; \; \; \text{(homopolymer)} \label{non-rand}
\ee
Similarly, $\langle (m - \langle m \rangle)^2 \rangle \sim (\fc -
F)^{-2}$ near the transition. The thermal fluctuations about $\langle m
\rangle$ are thus comparable to $\langle m \rangle$ itself.  The
divergence in (\ref{non-rand}) is analogous to the divergence in
interface height near a wetting transition.

\par
We now determine how results such as (\ref{non-rand}) are modified for
a random DNA sequence.  Sequence randomness is at
worst a marginal perturbation at the ($F=0$) melting transition in
3 dimensions~\cite{hwa,deniz,melt-refs}.  The application of a
Harris-like criterion~\cite{harris}, however, shows that the same
cannot be true for the unzipping transition: The typical variation per
monomer in the base-pairing energy of a pulled out section of length
$<m>$ scales like $<m>^{-1/2} \sim
\sqrt{\fc - F}$, which vanishes more slowly as the transition is
approached than the relevant energy difference $|\epsilon_0| - F^2
b^2/(2 d T) \sim \fc - F$.  To determine the correct critical behavior,
we focus on the free energy cost of pulling out a given monomer.
Define $\e(m)$ to be the free energy of a dsDNA molecule,
subject to an applied force $F$, of which exactly the first $m$
monomers are unzipped.  The change in \e\ from pulling out one
additional monomer should have the form
\be
\frac{d\e}{dm} = f + \eta(m) \; , \label{e-diff-eq}
\ee
which may be integrated twice to obtain the partition function $Z =
\int_0^{\infty} d\!m \exp[-\e(m)/T]$ .  Here $f$ is the average free
energy difference between an unzipped and a bound pair of
complementary monomers.  It vanishes like $\fc-F$ near the transition,
and reduces to the familiar $|\epsilon_0| - F^2 b^2/(2 d T)$ in the
absence of sequence randomness.  The additional term $\eta(m)$ takes
account of sequence-dependent deviations from the average; it reflects
the bare sequence (described by $\ett(m)$), dressed by thermal
fluctuations.  As long as $\ett(m)$ is a random variable with only
short-ranged correlations, it is reasonable to expect that $\eta(m)$
should also be short-range-correlated, with a correlation length on
the order of the typical size $\xi$ of the regions of local melting of
the dsDNA strand~\cite{xi-note}.  On long enough scales, we can then
take $\eta$ to be Gaussian white noise, with correlator
$\overline{\eta(m) \eta(m')} = \Delta \delta(m-m')$, where the overbar
indicates a ``disorder average'' over the possible realizations of the
quenched random base sequence.  The parameters $f$ and $\Delta$ may be
calculated from the $F=0$ partition function $Z_{\text{melt}}$, for
example in a low temperature expansion.  The model summarized
by Eq. (\ref{e-diff-eq}) can also be derived from a discrete,
Ising-like description of the dsDNA~\cite{ising}, and it still holds
both when the single strands that have been liberated are
characterized as freely-jointed or worm-like chains and
when there are significant excluded volume interactions~\cite{us}.

\par
The study of the unzipping transition can thus be reduced to that of a
single coordinate $m$ in the random potential $\e(m)$.  One
immediate consequence is that there is no large parameter that defines
a thermodynamic limit, and thus no equivalence between the ensemble
considered here and the conjugate ensemble in which $m$ is held
fixed.  Below the unzipping
transition, $f > 0$, and $\e(m)$ diverges with probability unity as $m
\rightarrow \infty$.  In the ensemble studied
here, the unzipping fork is thus
always confined to the vicinity of $m=0$.  In the absence of
randomness, the probability of unzipping $m$ monomers is $(f/T) \exp(-m
f/T)$, and one recovers,
e.g. (\ref{non-rand}).  If there is sequence randomness, the typical random contribution to $\e(m)$
is of order $\sqrt{\Delta m}$; the random part thus exceeds the average
contribution $f m$, which is responsible for the confinement, for $m
\lesssim \Delta/f^2$.  This length scale diverges faster than $1/f$ as
$f \rightarrow 0$, suggesting that a
typical value of $m$ might show a $1/f^2$ divergence instead of the
non-random $1/f$, with a crossover at $f \approx \Delta/T$.

\par
Because of the confinement near $m=0$, the
unzipping transition does not exhibit self-averaging (see below).
  Nonetheless, one can still calculate averaged quantities and
distributions over the possible realizations of randomness.  To do this, one wishes to study
 the disorder-averaged free energy $-T \overline{\ln Z}$.
The partition function $\zf(m) \equiv \int_0^m d\!m' \exp[-\e(m')/T]$ of a finite-sized system of $m$ (bound or liberated) monomers satisfies
\be
\frac{d \zf}{dm} = e^{-{\cal E}(m)/T} \; \; \; \text{and} \;\;\; \zf(0) = 0 \; ; \label{Z-diff-eq}
\ee
$Z$ follows simply by taking the limit of an infinitely long
polymer: $Z \equiv \lim_{m \rightarrow \infty} \zf(m)$.  Together, (\ref{e-diff-eq}) and
(\ref{Z-diff-eq}) form a system of coupled Langevin equations,
analogous, for example, to those describing the Brownian motion of a
massive particle, with \e\ playing the role of momentum and \zf\ that
of position.  The associated
Fokker-Planck equation for the joint distribution $P(\e,\zf;m)$ of \e\
and \zf\ at ``time'' $m$ follows in the usual manner~\cite{risken}:
\be
\frac{\partial P}{\partial m} = \left[ \frac{\Delta}{2}
\frac{\partial^2}{\partial \e^2} - f \frac{\partial}{\partial \e} - e^{-{\cal E}/T}
\frac{\partial }{\partial \zf} \right] P \; .
\label{fp-eqn}
\ee
By Laplace transforming with
respect to \zf\ and to $m$, one can solve (\ref{fp-eqn}) and obtain
an exact expression for the partial distribution $\int d\!\e P(\e,\zf;m \rightarrow
\infty)$ in terms of modified Bessel functions of order $2 f T/\Delta$; $-T \overline{\ln Z}$ and other thermodynamic quantities then
follow by integration.

\par
The device of
treating the quenched randomness as a Langevin noise thus leads to a
number of exact results.  In particular, the average
degree of opening $\overline{\langle m
\rangle} = - T \partial \overline{\ln Z}/\partial f$ satisfies
\bea
\overline{\langle m \rangle}
 & = &  \frac{2 T^2}{\Delta \Gamma(2 f T/\Delta)}
\int_0^{\infty} d\!y \, y^{2f T/\Delta -1} (\ln y)^2 e^{-y} \nonumber \\
&&- \frac{2 T^2 \Gamma'(2
f T/\Delta)^2}{\Delta \Gamma(2 f T/\Delta)^2} \; , \label{avg-m}
\eea
where $\Gamma'(z) = d \Gamma(z)/dz$.  The small $f$ behavior of
Eq. (\ref{avg-m}) is given by 
\be
\overline{\langle m \rangle} \simeq \frac{\Delta}{2 f^2} \sim (\fc
-F)^{-2}\, , \; \; \; \; \text{(random heteropolymer)} \label{avg-m-asympt}
\ee
confirming our expectations for a crossover
from a $1/f$ to a $1/f^2$ power law  when $f \sim \Delta/T$.  Similarly, the
singular part of the heat capacity associated with the unzipping
transition, $C \sim \partial^2 \overline{\ln Z}/\partial T^2$,
crosses over from a $1/f^2$ to a $1/f^3$ divergence.
One can also
compute the disorder-averaged values of higher cumulants of $m$.  For small $f$, $\overline{\langle m^2 \rangle - \langle m
\rangle^2} = T \partial^2 \overline{\ln Z}/\partial f^2 \sim 1/f^3$.
Unlike in the non-random case, the square root of this quantity
diverges more slowly than $\overline{\langle m \rangle} \sim 1/f^2$,
indicating that
thermal fluctuations in $m$ for a given realization of the quenched
randomness (and for thus a given heteropolymer) typically become small
compared to the mean value as the
transition is approached.

\par
A real space renormalization group approach to the model of
equations (\ref{e-diff-eq}) and (\ref{Z-diff-eq}) due to Le Doussal,
Monthus, and Fisher~\cite{dsf} gives further insight into the
unzipping transition.  This technique gives leading order results in the
limit $f \rightarrow 0$, where the authors have argued that it should be
exact.  In this limit, it allows one to calculate the
distribution $Q(\langle m \rangle)$ of thermal average values over
different realizations of randomness.  This takes the form of a
scaling function of $\langle m \rangle f^2/\Delta$:
\be
Q(\langle m \rangle) = \frac{f^2}{\pi \Delta} e^{-\frac{\langle m
\rangle f^2}{2 \Delta}} \int_0^{\infty}dw \, e^{-\frac{w \langle m \rangle
f^2}{2 \Delta}} \frac{\sqrt{w}}{w+1} \; . \label{Q-distrn}
\ee
This distribution yields the same asymptotic behavior of
$\overline{\langle m \rangle}$ near the unzipping transition as the
Fokker-Planck approach.  It also predicts that $\overline{(\langle m
\rangle - \overline{\langle m \rangle})^2} \sim 1/f^4$;
$\langle m \rangle$ for a polymer with a given random sequence of base pairs can thus deviate significantly from
the disorder average, and this system is not self-averaging.

\par
For the randomness-dominated critical properties reported here to be
observable, the variance $\Delta$ in the base-pairing energy must be
sufficiently large.  Then at the crossover from non-random to random
behavior, $f$ will also be large, and the typical value of $\langle m
\rangle \sim T^2/\Delta$ will be small enough that finite size effects
do not become an issue.  In this respect, dsDNA appears to be a very
good candidate system.  Under physiological conditions, the difference
in free energy of binding between polymers with only A-T base pairs and
those with only G-C base pairs is of order $T$, meaning that $\langle
m \rangle$ is only a few monomers when the crossover from pure to
random behavior occurs.

\par
In sum, we have described a randomness-dominated unzipping
transition of dsDNA, obtaining exact
expressions for the critical behavior and for the crossover from random to
non-random scaling.  Most notably, we find that when the base sequence
is random and has only short-ranged correlations,
the average degree of
opening $\overline{\langle m \rangle}$ diverges like $1/(\fc - F)^2$
as the pulling force $F$ approaches a critical value \fc, in marked
contrast to the $1/(\fc -F)$ divergence found when all of the base
pairs are identical.
It should be possible to arrive at analogous
results for the case of DNA whose base sequence has long-ranged
correlations (as may be the case for non-coding
DNA~\cite{dna-statistics}).  If a typical variation about the average energy
grows like $m^{\beta}$, then balancing this against $m f$ suggests
$\overline{\langle m \rangle} \sim 1/f^{1/(1-\beta)}$; the
short-range-correlated case is recovered when $\beta = 1/2$.
The biological
significance of our results remains to be determined:
Processes such as DNA replication and recombination often
involve unzipping of the dsDNA.  Usually, however, this is
accomplished by a molecular motor relying on an outside energy source,
so non-equilibrium effects must be considered.  More generally, the
dynamics of the unzipping transition is an open question.

\par
We have focussed on the case of unzipping DNA, but our results hold
equally well for a number of more conventional condensed matter
systems described by the Hamiltonian ${\cal H}_{\text{melt}} + {\cal
H}_{\text{pull}}$~\cite{us}.  The pulling of a Gaussian random {\em
hetero}polymer away from an adsorbing surface is a natural extension
of recent work on homopolymers~\cite{adsorbed}.  Other examples
include a heteropolymer under tension pinned to a bulk
defect~\cite{sommer}, a magnetic flux line in a type II superconductor
confined to a fragmented columnar pin and subject to a transverse
field~\cite{hatano,deniz}, and a simplified model of the corner wetting
transition in two dimensions~\cite{wetting}.  Related
models are likely to be relevant to the transverse surface
magnetization and surface specific heat near the Bose glass transition
of a bulk superconductor~\cite{hatano} and to adhesion
in a random environment~\cite{nagel}.

\par
After this work was submitted for publication, we learned that related
results had been obtained, in a different physical context, for a
discrete version of Eqns. (\ref{e-diff-eq}) and
(\ref{Z-diff-eq})~\cite{discrete}. It is a pleasure to thank
D. Branton, D.S. Fisher, and T. Hwa for helpful conversations and
T. Hwa for bringing~\cite{discrete} to our attention.  This research
was supported by the NSF through grant DMR97-14725 and through the
Harvard MRSEC via grant DMR98-09363.





\begin{references}
\bibitem{xi-note} Because $\xi$ diverges as the melting transition is
approached, our description thus breaks down very near this
transition, leading to effects that are beyond the scope of this
Letter.
\end{references}
\end{document}